# Statistical Considerations on Political Responsivity Analyzed Over a 70 Years Time Span

Matteo Cirillo

Dipartimento di Fisica and MINAS Lab

Università di Roma "Tor Vergata"

00133 Roma, Italy

## Abstract

The time evolution of the response of electors to political life is analyzed for the Italian Republic over a *70* years span in which *18* political elections have taken place. The basis for the performed analyses are the official data available from the Italian Ministry of Interiors exposing the results of the political elections from *1948* until *2018*. The attention is concentrated on parameters providing information on the responsivity of the electors to country's political life. These parameters, expressed in adequate percentages, are the effective number of voters and the percentage of these expressing blank or spoilt ballots. The time dependence of these parameters, over the analyzed period, shows regularities, correlations and interesting peculiarities. The analysis concerns the results for both Chamber of Deputies and Chamber of Senators available, for each election, all along the *70* years time span that are those relative to people voting on the national territory.

Analysis of votes expressed in national elections is an issue involving statistics, sociology, economics, and politics in a way that it is often not straightforward to achieve convergence and objectivity for the opinions. Beside the specific analyses an overall concern of all the political parties is the percentage of citizens expressing their vote. It has been often conjectured that the participation of citizens to the electoral process could decide the result of a specific election. The purpose of the present paper is to analyze the participation of voters to Italian elections and to analyze the percentage of those that expressed blank or spoilt ballots. We intend to show that few parameters related to these specific phenomena, when available for a substantial time span, can provide interesting information over the political sensitivity of a country ruled by democracy and free elections. The information that we analyze have been extracted from the public web site of the Italian Ministry of Interiors [1]. The data are organized for a technical analysis and we conclude commenting their impact and a possible sociological interpretation.

As anticipated in the abstract we perform our analysis for the two Italian Parliament chambers, namely Camera dei Deputati and Senato: from now on we will refer to these two chambers as *Camera* and *Senato*. From the available data the following parameters are extracted : the percentage of citizens effectively voting with respect the total number of electors having right to vote and this parameter is evaluated for both *Camera* and *Senato* and will be indicated respectively as $R_V(Camera)$ and $R_V(Senato)$. Then we introduce the parameters $R_B$ and $R_S$ which are respectively the ratio between the number of Blank and Spoilt ballots and the total number of effective voters: a parenthesis, as before, will indicate whether these parameters are relative to *Camera* or *Senato* . Finally we introduce the parameter $R_T = R_B+R_S$ as the sum of the percentages of Blank and Spoilt ballots.

In Fig. 1a we show the time dependence, from *1948* until *2018*, of the of the parameter $R_V$ for the *Senato*, which we there indicate as $R_V(Senato)$. We see in the figure that the dependence of $R_V$ on time has two slopes. The first part of the dependence, which covers the time span up to *1979*, is characterized essentially by a constant value with an intercept around *93%* indicating a relevant, and basically constant, participation/response of the voters and a noticeable will to express their opinion through the vote. After *1979*, however, we see that the parameter has a clear tendency to decrease: a linear fit of the data now reveals that a linear dependence (the linear correlation coefficient of the fitting is equal to *0.964* ) can reasonably be envisaged through the data. The slope of the line indicates that the number of citizens expressing their political will went down, after *1979* of a rough *2%* every *5* years  (the constitutional duration of an elected parliament in Italy). Beside the adequacy of the linear fitting, the overall tendency of $R_V$ *(Senato)* to decrease is evident. The reasons for this phenomenon has been investigated in other publications [2], but the interesting thing that we wish to note in Fig. 1a is that the time response is essentially linear with just one change in derivative. We will go back to this aspect later after commenting more data.

A plot analogous to the one shown in Fig. 1a can be obtained for the parameter $R_V$ *(Camera)*, namely the percentage of people voting with respect to total number having right for the Camera dei Deputati. Indeed results similar (dated up to *2008*) to those shown in Fig. 1a, were reported for the Camera in ref. 2 (see Fig. 1 of that publication).  Plotting the results for $R_V$*(Senato)* and $R_V$*(Camera)* on two Cartesian  axes with linear scales for the whole analyzed time scan, as shown in Fig. 1b,  we realize that there is no difference between the two. The data lie on the bisector line of the quadrant meaning that the two set of data express identical information; the linear correlation coefficient for the straight line is in this case differs from 1 for less than *0.1%* meaning that we can reasonably assume  $R_V$*(Senato)*$\cong R_V$*(Camera)*.

In Italy the Constitution allows to vote for the Chamber of Senators only people *25* years old and above; we can see that, as far as the will to express the vote is concerned, this age constriction does not make any appreciable difference with respect to the Chamber of Deputies, where all people above the age of *18* years can vote. The fact that $R_V$ is the same for both *Camera* and *Senato* is likely related to the limited percentage of electors who can make the difference between the two Chambers:

these are just those with the age in the interval *(18-25)* with respect to the total numbers of voters at the time of a given election.

In Fig. 2a we show, for the *Camera*, a plot where we put $R_B$ on the horizontal axis and $R_S$ on the vertical axis; we can see here that the data, except for the circled point [3], fall mostly on two parallel lines whose slope is equal respectively to *2.04 ± 0.12* and *2.15 ± 0.05* and intercepts *1.04 ± 0.20* and *-0.55 ± 0.10* . The linear correlation coefficient of the two lines are respectively *0.986* (upper line) and *0.998* for the lower line. Thus, the two slopes are consistent within the errors and their weighted average is *2.11 ± 0.05*  meaning that, the percentage of spoilt ballots can be obtained, with a rough *4%* discrepancy, from percentage of blank ballot by multiplying a factor two. We note that all the *7* points fitted by the lower line are results of elections up to 1994 (included) while all the results after 1994 lie in the upper curve. Apart for the specific numerical proportionality constant, this plot shows that the two parameters are linearly correlated: the overall amount of spoilt ballots does not correspond to random mistakes made by the electors, but it is related to the same intention expressed by the blank ballots to not vote for any party. It is known in fact that most of the spoilt ballots contain just scribblings and graphic fantasies indicating, more than a mistake, the will to not vote.  A plot very similar to the one of Fig. 2b can be obtained for the *Senato*.

From what we said in the previous paragraph we argue that it is reasonable to consider the sum $R_T = R_S+R_B$  as an indicator of the intention of the voters to not express a vote in favor of any political party in a given election. In Fig. 2b we show the time dependence of the parameter $R_T$ *(Camera)* over the whole time span investigated where an overall rise of the parameter in time is evident.  Although the behavior does not seem regular at a first sight, we see in Fig. 2b that the data of four elections are very well intercepted by a straight line and the other points are so well distributed below and above this line that the linear correlation coefficient of the fit is essentially the unity.  We can see, however, that the points circled by the oval in the plot, in the low-right part, are clearly far from the tendency to increase in time of $R_T$. These points correspond to the data relative to elections of *2006, 2008, 2013, 2018*. The dashed arrowed line in the figure is a guide to the eye evidencing the abrupt factor *2.8* decrease of the $R_T$ occurred between the election of *2001* and those of *2006* whose interpretation is not easy. One could expect $R_T$ tending in time to a sort of plateau value, assuming that in a sane democracy this parameter cannot grow forever, but this expectation would be somewhat validated by a more smooth temporal variation and not by a sharp, and substantial, decrease like that shown in Fig. 2b. Even admitting that the election of *2001* $R_T$ could be affected by an unfortunate casuality [3], it is unavoidable to admit from *2006* onward a substantially different "tendency".

Another interesting information comes from the plot of Fig. 3 which further clarifies the "spurious" nature of four points of Fig. 2b. In this figure we report on the horizontal axis the parameter $R_T$ *(Camera)* while on the vertical axis we report the parameter $R_V$ *(Camera)*; recall that $R_V$ is the number of people expressing their will in an election normalized to the total number of individuals having right to vote. The straight line we see between the data is a linear fit which returns a negative slope equal to *-0.015±0.002* and an intercept equal to *99.7± 1.4*. The linear correlation coefficient of the line returned by the statistical fit is *0.912*. The red points inside the oval contour correspond to the same ones inside the oval in Fig. 2b; even we see here that these points are far from the somewhat regular dependence of the other election points. Fig. 3 is interesting because it shows that the "historical" tendency is that more people participate to the elections the less is the amount of invalid votes: in other words an increased percentage of voters indicates an increased tendency to express a valid vote. Thus, the conjecture that an increase of the number of voting electors increases the probability of getting valid votes makes sense, but this is an overall result which does not relate the increase to any specific politic direction, it is just an overall tendency.

The straight line in Fig. 3 can be interpreted as a politics "approval-rating line" and the interesting thing is that the intercept to zero of this line is *100%*, within a *1.4 %* error, meaning that if all the electors having right were to vote the percentage of "unexpressed" votes would tend to zero. It would be definitely interesting to analyze similar data for other countries. From what we just said, however, it is clear that the points inside the oval in Fig. 3 (the same in the oval in Fig. 2b) represent and evident deviation from the linear tendency: these point, in Fig. 3 correspond to a low $R_V$ and a low $R_T$ which are far from the "approval-rating line" (representing a *50* year tendency).

Overall, the statistical analysis herein reported is at undergraduate level and the straight lines interpolating the data are just least square fittings. All the linear fittings, however, have correlation coefficients above *0.9* meaning that, considered that in the worst case the number of data generating the fit was *7*, the probability of the considered data not being linearly correlated was less than *0.05%* [4]. This result represents a reasonable degree of confidence for the presented analysis. It is remarkable that complex socio-political phenomena analyzed over a substantial time span display such simple dependencies. However, this it is not true in general and we can realize it in Fig. 1a, Fig. 2b and Fig. 3. In Fig. 1a we observe a significant discontinuity in the slope of the straight lines beginning after *1979*, while in Fig. 2b and Fig. 3 we see that an abrupt transition taking place in *2006* takes the response far away from what one could expect from a *50* years "regular" tendency. The general behavior of natural phenomena is nonlinear and the reduction/approximation to linear responses often provides just a first approach in taming problems, and often for slight intervals of the independent variables. Here instead we see a scenario a bit different for the response of a percentage of voters : it is mostly linear with some discontinuities/irregularities.

Before continuing with the analysis we believe it is better to step from the "normalized" values we gave in our figures to rough, unnormalized data. The plot of Fig. 1 tells us that, from 1979 until 2018 roughly 10 millions of Italians have decided to not vote while the plot of Fig. 2b and Fig. 3 tells us that between the elections of 2001 and 2006 the number of blank and spoilt ballots went down of about three millions. We believe it makes sense understanding of the phenomena generating those effects.

It would be interesting to justify the objective anomalies from the perspective of the socio-political mass phenomena. The conclusion that we extract from the data is that overall, simply ruled, relations exist, but also "singular" points in which we see the response of the percentages of voters we are interested in changes abruptly. The only "technical" treatise we are aware on for of the onset of discontinuities in behavioral science, decision making, biology and other disciplines was put forward by Renée Thom [5]. Thom's analysis was concerned with a general model for the singularities of nonlinear dynamical models, and related potentials. Excellent sources of information and references for this model can also be found in refs. 6 and 7. Within Thom's analyses an abrupt transitions in behavior is caused by the background existence of different stable states. He classified seven of these kind of abrupt discontinuities baptizing these as "catastrophes". We can say that the stepping observed in Fig. 1a from a very high, almost "constant", participation of citizens to the voting process, to an attitude expressing a contantly decreasing partecipation is reminiscent of the sharp transitions due to the folding of behavioral surfaces described in catastrophe theory [8]. In this specific case the two stable states could be "partecipation" and "indifference", both in relation to political life. In these terms it is clear that from *1979* on "indifference" prevails generating a constant decrease of people participating to the elections. This is, however, all we can say since framing, technically, the results presented in terms of the *7* catastrophes models would be hardly achievable at this point.

We judge that an abrupt change in behavior is also what occurs in Fig. 2b. While Fig. 1 concerns the electors who decide to not vote, namely to not express their opinion at all, Fig. 2b concerns the percentage of the electors who, voting, decided to make clear the intention to not like any political party (by a blank or spoilt ballot). Even in this case we see an abrupt transition from a tendency to increase constantly in time of $R_T$, whose duration is about *50* years, to substantially lower values observed in the last four elections. We have also seen that this recent tendency, generates an abrupt change in another parameter plane: in particular, the plane *($R_T$, $R_V$)* of Fig. 3 shows a noticeable regularity of the relation between these two parameters for almost *60* years but, while the transition that started in *1979* did not affect the relation between these two parameters, the transition of *2006* did. Thus, what we can say is that the four "anomalous" points of Fig. 2b and Fig. 3 is an unprecedented result of political expression and it would be surely interesting to further investigate what the causes for the phenomenon are.

It is also worth mentioning that, apart for the discontinuity occurring after *1979* in Fig. 1a, the percentage of electors voting $R_V$ has proceeded along the negative slope until *2018*. Thus, the discontinuity and abrupt change of Fig. 2b cannot related to the overall percentage of electors deciding to vote since this parameter has kept the same tendency. The "spurious points in the ovals of Fig. 2b and Fig. 3 are evidence of an "internal" structure inside the parameter $R_T$ which is not related to the "regular" overall decrease $R_V$.

As we said above a "technical" explanation of the results that we have presented would surely not be easy. By technical explanation we mean finding a theory which could provide a mathematical fit for the data. Catastrophe theory was worked out to technically describe critical phenomena and discontinuities, but the potentials describing the models should be figured out and that would surely not be straightforward in our case. We must say, however, that the kind of discontinuities illustrated in Fig. 1-3 are also reminiscent of phenomena associated with phase-transitions in physics and chemistry [8]. It is very typical of thermodynamic quantities to exhibit discontinuities in presence of phase transitions and indeed phase transitions are classified on the basis of the discontinuities of specific potentials [8] :even here, as before, although our data are reminiscent of physical phenomena related to phase transitions and related discontinuities, a technical analysis would be too qualitative to be really significant.

Leaving apart now possible mathematical fitting of the plots we presented and the noticeable analogies with phase transition and catastrophe models in physics and other sciences, let us make instead a short review the socio-political phenomena that could motivate stable changes like the one shown in Fig. 1a and the consequent growing indifference toward voting in elections.

Up to 1979 political life in Italy was mainly a matter of government and opposition between the Cristian Democrats party (Democrazia Cristiana, DC) and the Comunist Party (Partito Comunista Italiano, PCI). The DC had been always pivoting governments since 1948; along the 70s, however, the percentages of votes for both DC and PCI were both of the order of 30%. Following a season of extreme left and right wing terrorism DC, PCI, and other parties, started collaborating in order to keep the nation away from social disorders, dangers for democratic life, and financial bankruptcies. This collaboration led in 1976 to the "governo di solidarietà nazionale" (government of national solidarity) headed by Giulio Andreotti (DC) which was "externally" supported by PCI, although no communist deputies or senators were appointed ministers and/or were bearing direct government responsibilities.

Following the "solidarity" at national level collaboration started level at "periferic" levels between DC and PCI. Somehow the tendency was initially motivated by Enrico Berlinguer (secretary of the PCI) idea of "Compromesso Storico" (Historical Compromise), a program aiming to give rise to a government of the country based on the substantial (and formal) participation of both DC and PCI. All along the 80s the PCI joined DC and other center or left wing parties, namely Socialist party (PSI), Social Democrat Party (PSDI) and Republican Party (PRI) to give rise to local governments all over the country.

In 1992 an investigation (Mani Pulite) started in the Milan cleared out that all over Italy, not only in Milan, political parties, more than cooperating on the solution of social and economics issues, were mainly caring for support of their activity through illegal funds rise and corruption in public management. It turned out that the problem of parties exercising control, and carving up, on all kind of management of public resources, had become a dominant attitude all over the country: way ahead of 1992, this problem indeed was well present in citizens consciousness. Thus, a general sensation of impotence grew all over the 80s toward the changes that political elections could generate and it is likely that the decreasing interest, shown in Fig. 1a, for voting was generated by a lack of credibility of the political class, a political class that was mostly flushed out by investigations and court cases. That year, 1992, is acknowledged as the end of the First Italian Republic (started in 1948).

The new political party "Forza Italia" which merged from the "Mani Pulite" flushing relied, on a new generation of deputies and senators, and the assumption of these not being involved in the previous political management was a relevant point submitted to the public opinion. Moreover, the fact that the two main allied of Forza Italia, the right wing party Alleanza Nazionale (former Movimento Sociale Italiano) and the the "Lega Nord" never had management responsibilities in national government - up to 1994 - was also considered a relevant point to submit to the attention of voters. In 1994 the coalition of these three parties won the national political elections, but, as we see in Fig. 1, that was not enough to stop the increasing lack of interest for voting of Italians that has continued to date. If we associate disorder to lack of knowledge of voters intentions, it really looks like, after 1979, the country had a transition to a state in which disorder, and entropy, constantly increases.

All along the "seconda repubblica" years the impression that the pre-1992 tendency to power-sharing between political parties persisted: in spite of TV talk shows and programs in which exponents of political parties did attack each others, the facts did not seem to provide evidence of substantial democratic debates/controversies. The most known episode that political parties, appearently opposing each others, were indeed still hosting insane cooperation to share resources, was the "Mafia Capitale" investigation which started in 2010 in the city of Rome. In general, for several laws it is not really clear what the role of the oppositions has been in balancing the effects, or at least in pointing out clearly aspects of the laws being promulgated. Other episodes of corruption unfortunately involved, at the highest levels, even those political parties (Alleanza Nazionale, Lega Nord) that had the possibility, after 1994, to be involved in relevant government responsibilities.

After 2009 indeed a new political formation was founded, the "Movimento Cinque Stelle" (Five Stars Movement): the declared motivation for the birth of this party was, again, fighting the insane transversal agreements of existing political parties and working on issues that could effectively interest people. Whether the purposes of this party, receiving the relative majority of votes in 2018, will be effective in the long range to generate more credibility and transparency of thew political actions it is perhaps too early to say.  In any case, even in the 2018 elections the number of electors voting has kept decreasing: the phenomenon that really makes us to think in terms of catastrophes phenomenology/theory is the fact that once a transition occurred then the system has remained, in a stable manner, in the specific state (growing indifference toward politics) in which it collapsed.

Now, beside the fact that, since 1992, judges and investigations seem to have become elements having a relevant weight in addressing Italian politics, let us go back to the linear decrease of participation of electors to the votes and its possible motivations. We list below the main "macroscopic" events following political elections since 1994, meaning by "macroscopic" those happenings in the political life which have been objectively visible by electors.

In 1994 the coalition headed by "Forza Italia" won the elections, but, within one year  the coalition collapsed and a government headed by Lamberto Dini drove the country to new elections in 1996. In 1996 the "Ulivo" coalition won the elections but, in 1998 the government collapsed due to somewhat demagogical motivations brought up by the extreme left wing. Whatever the motivations were, effects could be judged afterwards: the head of the "Ulivo" coalition, Romano Prodi was appointed as head of the European Commission and Massimo D'Alema, a leader of the major political parties of the coalition (Partito Democratico della Sinistra) was appointed prime minister. D'Alema government stayed in charge for little more than one year and then left space to another "institutional" government headed by Giuliano Amato.

Objective divisions and internal controversies inside the Ulivo  led it to lose the 2001 elections in favor again of the right wing coalition headed by "Forza Italia" and his leader Silvio Berlusconi. The (2001-2005) government of the right wing was successful to stay in charge for 5 years; beside the internal divisions of the coalition which led to replacement of relevant ministers, the main activity of this government was not pivoted on the issues invoked during the electoral campaign, namely simplification (and transparency) of public administration, public healthcare, taxation, and promoting private investments. This government had an intense activity to promulgate laws which were, objectively, not in the interest of the majority of the electors. All the problems that Italians had to suffer after 2002, due to the stepping from lira to euro and the consequent substantial loss of spending power of salaries, were just presented as a responsibility of the previous government but no actions/programs were attempted to help people. In those years the slaughtering of Italian culture, public instruction and research started, and never ended since then. The political action in culture and education has become an issue for propaganda: it is difficult to keep track of all the changes and the new laws introduced in public schools since 2001, by left and right wing governments. Naturally, all changes and laws have been thought to need a minimal (if not zero) financial support. Getting degrees in high schools has become a moving target with disciplines programs changing continuously and new rules appearing as necessary corrections of the previous ones. This instability has generated cultural deficiencies in new generations but, most importantly, lack of interest for knowledge and science. The only way to look at this process, and be positive, is just speculating that laws to come can be worse.

The "Prodi" coalition won again the elections in 2006 (the first elections in which the anomalous data of Fig. 2b and Fig. 3 showed up), however, few months after the elections, the winning coalition was rapidly set in troubles by the creation of a new political subject (Partito Democratico) by Prodi and collaborators. It is remarkable that this political "creation" could very clearly be foreseen as a severe threat for the heterogeneous coalition, so clearly that it appeared like a deliberate act to generate a collapse. Since then nobody has ever explained, with trustable arguments, what the necessity was for creating, few months after the elections, a new political party. At that time government issues should be seriously considered and the commitments of the electoral campaign be respected. The creation of the Partito Democratico indeed was not an issue in the program Prodi had presented as candidate: this program was a 281 pages document but contained absolutely no traces or promises that the elections would lead, as a primary objective, to the creation of a new political subject. The compactness of the "left" wing has been an issue for a long time in Italy but the creation of the Partito Democratico, in 2006/2007 surely did not represent a solution, neither for the 2006 Prodi Government (collapsed in 2007) nor in the long range.

The fourth Berlusconi government took birth in 2008. It was made evident to the electors, after three years, that this government was leading the country to bankrupt. What really was the degree of certainty of this danger it is hard to tell objectively, but it is sure that the "technical" government (headed by Mario Monti) that was put in charge to take care of the presumably disastrous situation imposed rather severe financial efforts and sacrifices to Italians.

In 2013 a slight majority led the left wing coalition, headed by Pierluigi Bersani (secretary of the Partito Democratico), to win the elections. Nevertheless, Bersani was not successful in setting up a government and, along the five years 2013-2018, three governments, headed respectively by Enrico Letta (April 2013-February 2014), Matteo Renzi (February 2014-December 2016), and Paolo Gentiloni (December 2016-June 2018), ruled the country. Although, according to the constitution, inside an elected parliament attempts can be made to provide a national government, it is clear that the indifference of electors deciding not to vote can be understood in consideration of the fact that the issues of electoral campaigns are often not followed by corresponding government activity. Indeed, the same individuals leading the election coalitions often disappear from the political scene after the elections, along with their promises and programs.

What happened after the 2018 elections fits well along the line described in the past paragraph: the Movimento Cinque Stelle, after getting the relative majority of votes, has been responsible for generating two governments first with right wing allied (mainly Lega Nord ) and then with left wing allied (mainly Partito Democratico). Both Lega Nord and Partito Democtratico had been severely opposed and criticized by the "Movimento" in the past decade.

From the short historical survey given above we would say that the constant decrease of interest for electors to vote shown in Fig. 1 can be, if not justified, at least somewhat understood and we would say that no substantial and trustable attempts have been made on the side of the politics in trying to revert the tendency. As we said before, efforts attempting to provide a government team inside an elected parliament is a constitutional commitment of the President of the Republic, however, it is a fact that the governments in charge have not represented in the majority of the cases, the issues, and the teams, the electors voted for. The growing tendency to not vote has a noticeable 40 years stability and this fact becomes worrisome now if we consider the data of Fig. 2 and Fig. 3: here we have no comments and observations in terms of facts and socio-political issues for interpreting the data, but it would be important to clarify the anomalies epitomized in those plots. Apparently, starting in 2006 a "state" developed for which the subset of electors voting blank or spoilt ballots has become rather different with respect to the past. The phenomenon, generated abruptly (in 2006) seems to have a certain degree of stability since then: understanding its nature, generating a mis-alignment with a 60 years tendency, would be more than desirable, almost necessary indeed.

It was surely not our intention to provide negative impressions of a nation where millions of individual are productive, generous, intellectually active and love their country art, culture, and environment. It is for this reason indeed that the political class should be more sensitive to the interface with citizens and work on the gap that keeps increasing between politics and relevant needs and interests of the people. Italians are getting aware of direct or indirect misleading TV propaganda and media manipulation and it is a fact that the organization/growth of the "Movimento Cinque Stelle" in the past ten years was managed mainly by the web on a specifically engineered platform. For long time leaders of the "Movimento" avoided systematically to appear in TVs, or even being interviewed, knowing that this would be appreciated by their supporters. It is time for politics to work hard on transparent and efficient public administration which is still suffering from the long wave of insane corruption established by the Fascist government. The attempt made in the past two decades to digitalize public administration, a relevant issue nowadays, has turned into management nightmares and episodes of corruption have been investigated in relation to the activity of companies handling the software platforms. Italians are paying, for each liter of gas bought for transportation, a bit less than half of it for taxes: this is justified by natural terrible disasters occurred in the country and severe financial commitments of the past 100 years. An Italian economy Minister said that paying taxes should a great pleasure because it concerns helping people with welfare, improving services in the country and else, true : however, it should also be made clear, and readily accessible to citizens, how is all the taxation income spent.

**FIGURE CAPTIONS**

Figure 1. (a) Time dependence of the ratio $R_V$ between the number of voters and the total of all the citizens having right to vote. The negative slope of the straight line indicates a rough *2%* decrease every 5 years, the time duration of a Parliament in charge in Italian Constitution. (b) The same ratio as above for Camera and Senato on the Cartesian axes shows that there is no distinction between the two percentages which lie on a straight line.

Figure 2. (a) The same kind of plot like in Fig. 1b where we report now on one axis the percentage of blank ballots and on the vertical the spoilt ballots. The slopes of the two straight lines, which give roughly *2* as an average, tell us that for every blank ballot there are *2* spoilt ones. The linear correlation between $R_B$ and $R_S$ tells us that these correspond somehow to the same intention to not express a vote. (b) The time dependence of the sum of the percentages of blank and spoilt ballots $R_T$ as a function of time. Even here we can see a grossly linear dependence (see text for fitting parameters of the shown straight line). However, the red points in the oval are far from the linear growth tendency.

Figure 3. The correlation between the percentage of voters at the *Camera* and the percentage of these responding with blank and spoilt ballots. The four points inside the oval perimeter correspond to the same shown in Fig. 2b in the "time domain". We define the straight line as a politics "approval-rating" line.

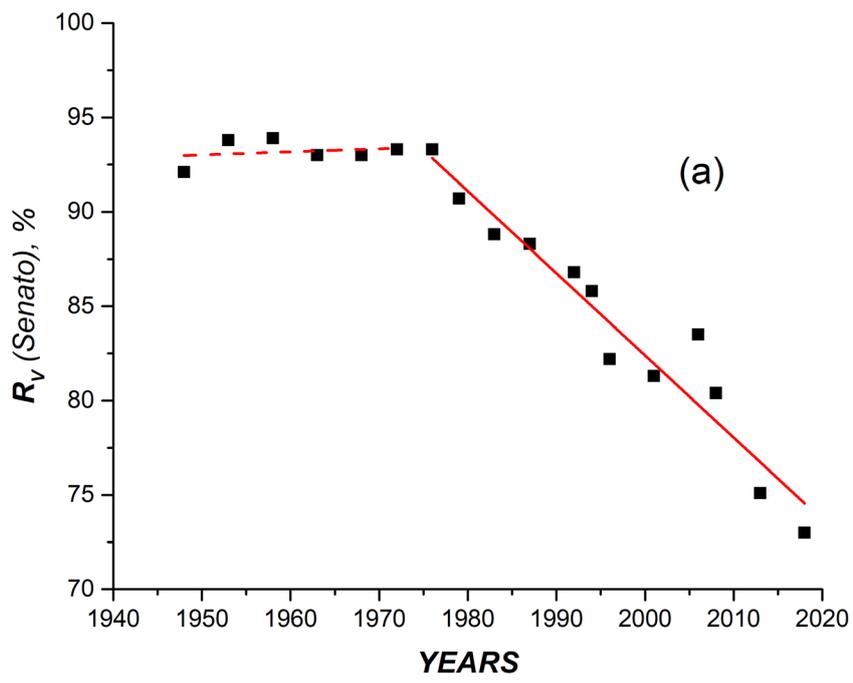

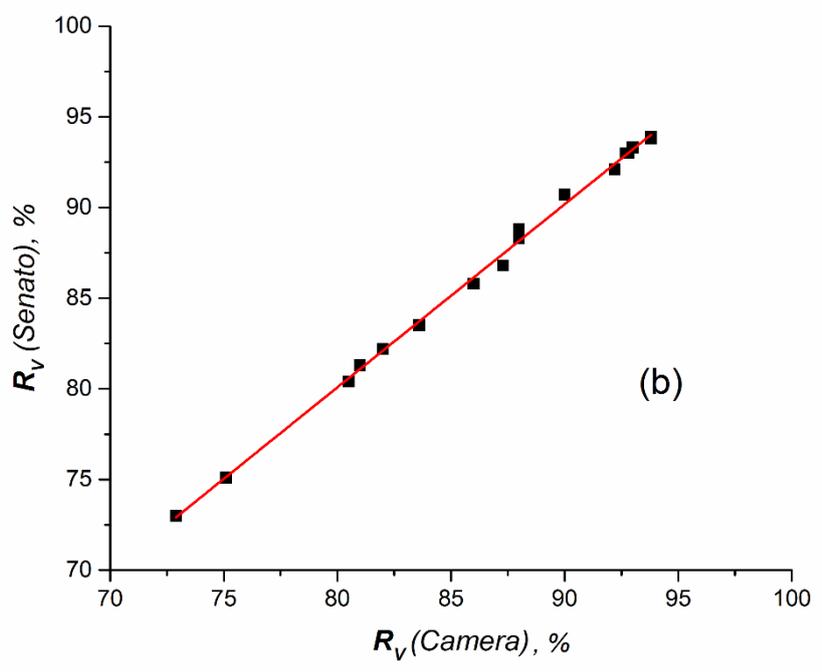

Figure 1, M. Cirillo

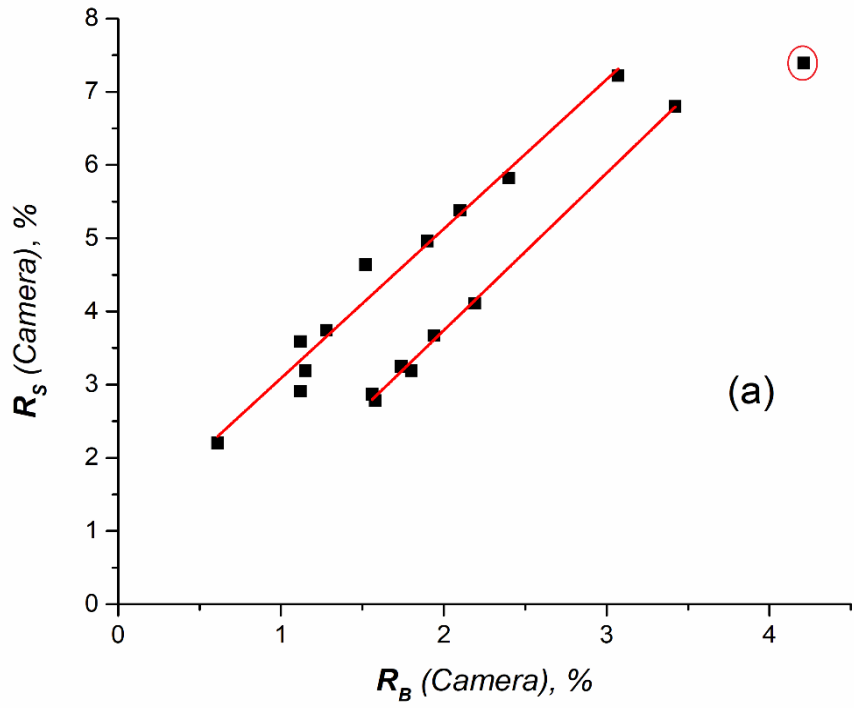

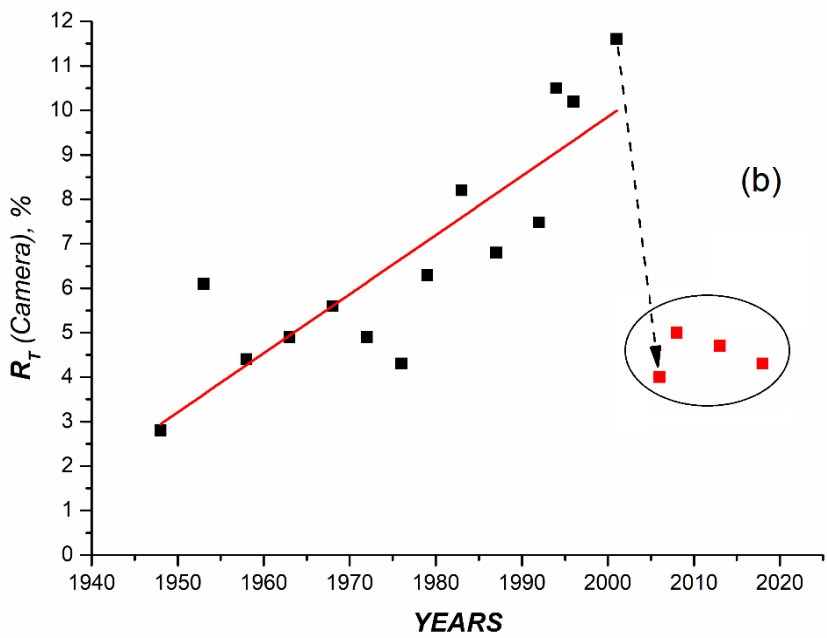

Figure 2, M. Cirillo

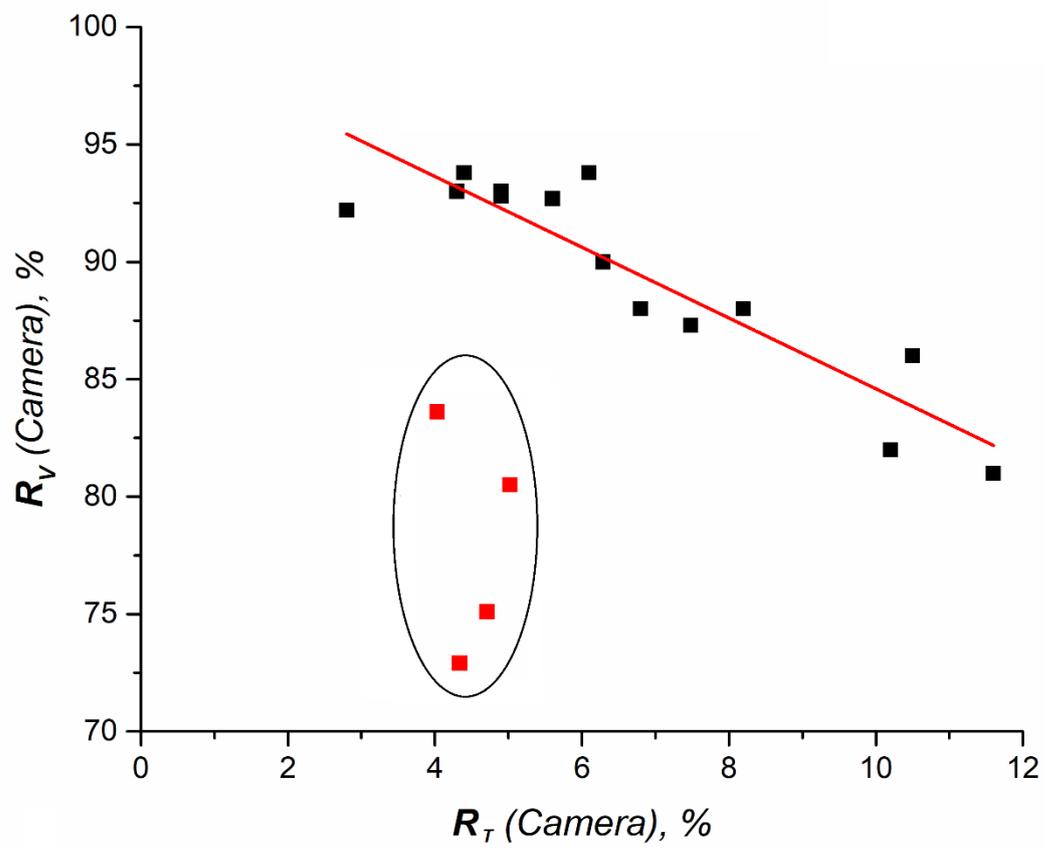

Figure 3, M. Cirillo